**Efficiency enhancement of a harmonic lasing free-electron laser**

E. Salehi[1], B. Maraghechi[1,a)], and N. S. Mirian[2]

[1]Department of Physics, Amirkabir University of Technology, 15875-4413 Tehran, Iran

[2]School of Particle and Accelerator Physics, Institute for Research in Fundamental Sciences (IPM), 19395-5531 Tehran, Iran

The harmonic lasing free-electron laser amplifier, in which two wigglers is employed in order for the fundamental resonance of the second wiggler to coincide with the third harmonic of the first wiggler to generate ultraviolet radiation, is studied. A set of coupled nonlinear first-order differential equations describing the nonlinear evolution of the system, for a long electron bunch, is solved numerically by CYRUS code. Solutions for the non-averaged and averaged equations are compared. Remarkable agreement is found between the averaged and non-averaged simulation for the evolution of the third harmonic. Thermal effects in the form of longitudinal velocity spread are also investigated. For efficiency enhancement, the second wiggler field is set to decrease linearly and nonlinearly at the point where the radiation of the third harmonic saturates. The optimum starting point and the slope of the tapering of the amplitude of the wiggler are found by a successive run of the code. It is found that tapering can increase the saturated power of the third harmonic considerably. In order to reduce the length of the wiggler, the prebunched electron beam is considered.

**I: INTRODUCTION**

a)behrouz@aut.ac.ir

X-ray never stops revolutionizing the understanding of matters, and creating new sciences and technologies. High-gain free-electron laser (FEL) amplifiers hold great prospects as high power, coherent, and tunable radiation in the x-ray regions of the electromagnetic spectrum. Utilizing nonlinear harmonic generation when bunching the harmonics is driven by the fundamental frequency in the vicinity of saturation is a possible way for obtaining x-ray wavelengths.[1-7]

Planar wiggler FELs allow resonant interactions with the odd harmonics of the fundamental resonant radiation field. However, intensity of the harmonics is rather small. For example, the third one is typically at the level of a percent of the fundamental intensity,[1,3-5,7] and the higher harmonics are much weaker. To amplify the harmonics, various seeded FEL schemes [ 8-18 ] were proposed and intensively studied around the world. High-gain harmonic generation (HGHG)[8-12], is one of the harmonic amplification FEL schemes, where the seed in a first undulator (modulator) is used to induce an energy modulation in the electron beam, and convert the energy modulation into density modulation using a magnetic chicane. Then this bunched beam emits amplified radiation in a following undulator (radiator) tuned at one of the harmonic frequency of the modulator. Ref. 11presented the first demonstration of the generation of a supperradiant pulse in the long radiator of a single stage cascade FEL, by seeding the undulator with an external laser. The first operation of the high-gain harmonic generation FEL, seeded with harmonic generated gas, is reported in Ref. 12. The second technique is called harmonic optical klystron (HOK) that has the same configuration as HGHG. In the HGHG the optical power grows exponentially while in the HOK the optical power grows quardratically. In Refs.13-15, cascade harmonic optical klystron are discussed. The up-frequency conversion efficiency for HGHG scheme is relatively low, so the echo-enabled harmonic generation scheme is proposed and studied for the generation of high harmonics using the beam echo effect[16,17]. This method uses two undulators, two dispersion sections, and one radiator, and has a remarkable up-conversion efficiency and allows for the generation of high harmonics with a relatively small energy modulation. Recently, McNeil *et al.*[18] proposed a harmonic lasing FEL amplifier in a one-dimensional limit that can be extended to higher harmonics by suppressing the interaction at the fundamental resonance while allowing the harmonics to evolve to saturation. They showed that this configuration can be used in the extreme ultraviolet and x-ray regions of the spectrum.

Two novel methods to suppress the interaction at the fundamental resonance without affecting the third harmonic lasing were outlined in Ref. 18. In the first method, a series of relative phase changes between the electrons and pondermotive potentials of the resonant fields at a series of points along the FEL interaction are used. In the second method, two different settings for the undulator are considered by changing the wiggler magnetic field while keeping the wiggler period, $\lambda_w$, and the initial average electron beam energy, $\gamma$, constant (see Fig. 1). Recently, it has been shown that the first method is inefficient in the case of a self-amplified spontaneous emission (SASE) FEL.[19] They suggested a modification of a phase shifter method which can work in the case of a SASE FEL.

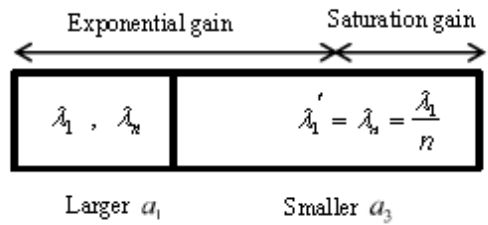

Fig. 1. Schematic representation of an harmonic lasing with two different settings

The thermal effect of the electron beam on the harmonic gain is particularly important. The kinetic theory has shown that the thermal FEL lowers the growth rate.[20,21] The energy spread is of particular importance in short wavelength FELs, because they are more sensitive to thermal effects.[22,23] It was concluded in Ref. 18 that harmonic lasing at $\lambda_3$, in the first method, is more sensitive to the emittance and the energy spread than the fundamental lasing at $\lambda_3$ in the second method.

The intrinsic efficiency of the FEL is low. By increasing the energy of the electron beam, the efficiency reduces further. Therefore, for the x-ray FEL, efficiency is very low. For this reason, much attention has been given in the literature to schemes for the FEL efficiency enhancement.[24-28] In Ref. 25, in order to increase the efficiency the phase velocity of the electromagnetic radiation is reduced by increasing the wave number of the wiggler after the saturation point. In Ref. 27, electrons are accelerated in the wiggler by the injection of rf power. An increase in efficiency is demonstrated experimentally in Ref. 28 when the electron beam energy is detuned from the resonance condition. Recently, it was experimentally shown that the efficiency was increased by reducing the amplitude of the wiggler after the saturation point for both the fundamental resonance wavelength and the third harmonic.[29]

In order to reduce the length of the FEL, a prebunched electron beam is used.[30-33] In Ref. 34, a nonlocal theory has been used to investigate the effect of beam prebunching on the gain and efficiency. The transient analysis of a prebunched electron beam was studied using the Green function method in Ref. 30, and the main concern of Ref. 31 was to devote to the one-dimensional treatment of the high gain Compton regime. A prebunched FEL in the Compton regime is simulated in Ref. 32.

The purpose of the present study is to use the concept of the linear and nonlinear tapering of the wiggler field to increase the efficiency of the harmonic lasing FEL. To this end, by decreasing the wiggler amplitude linearly or nonlinearly at the saturation point, the resonance condition of the FEL is restored, which will result in higher intensity UV radiation. The slippage of the radiation with respect to the long electron bunch is ignored. In the steady-state simulation of the FEL, it is assumed that each segment or beam slice interacts with the electromagnetic field in an identical manner so that the simulation treats only one slice. Equations describing harmonic lasing FEL are derived. This set of equations is solved numerically using CYRUS 1D code in one-dimension.[35] CYRUS, which was developed by N. S. Mirian *et al*.[35] to study nonlinear harmonic generation. This code like MEDUSA[36,37] employs nonaverage equations. The third harmonic lasing is considered so that the operating wavelength is in the ultraviolet (UV) domain. Second, we compare two different simulation codes, one of which uses wiggler-averaged orbit dynamics[2] and the other (CYRUS) does not. These codes deal very differently with almost all aspects of the FEL formulation included. Also, we investigate the effect of energy spread and prebunching of the electron beam on the harmonic lasing FEL. Evolution of the thermal distribution function, Kolmogorov entropy is studied.

The organization of the paper is as follows. We describe the one dimensional (1D) formulation used in the non-wiggler-averaged-orbit approximation for harmonic lasing FEL in Sec. II. In Sec. III, simulations are performed to compare the two different codes, investigate the thermal effects, increase the efficiency using linearly and nonlinearly tapered wiggler, and compare the different taperings and study the effects of prebunching in harmonic lasing FEL. A summary and discussion are presented in Sec. IV.

**II: BASIC EQUATIONS**

The numerical simulation of the harmonic lasing is conducted using the CYRUS 1D code that is written in standard FORTRAN90. The one-dimensional formulation represented in MEDUSA 1D[36] is a subset of the complete three-dimensional formulation of MEDUSA,[38,39] that includes time dependency and harmonics. The formulation treats the planar wiggler model and the radiation field is represented as a superposition of Gaussian modes.[38] The vector potential of the radiation field, in plane-polarized form, is

$$\delta \mathbf{A}(z,t) = \hat{e}_x \sum [\delta A_h^{(1)} \sin \varphi_h + \delta A_h^{(2)} \cos \varphi_h], \quad (1)$$

where $\delta A_h^{(i)}$ with $i = 1,2$ are the amplitudes that are assumed to vary slowly in $z$ and $t$ and $h = 1,3,5,...$ denotes the harmonic number, $\alpha_h = h(k_0 z - \omega t)$ is the phase of the $h$ th harmonic of the angular frequency $\omega$ with $k_0 = \omega/c$. We try to go from UV wavelength to X-ray wavelength, so the collective Raman effects are negligible. Averaging Maxwell's equations over the time scale $2\pi/\omega$, the field equations take the form

$$\left(\frac{\partial}{\partial z}\right)\begin{pmatrix}\delta a_h^{(1)} \\ \delta a_h^{(2)}\end{pmatrix} = \frac{\omega_b^2}{2h\omega c}\begin{pmatrix}\left\langle \frac{u_x}{|u_z|}\cos\alpha_h \right\rangle \\ -\left\langle \frac{u_x}{|u_z|}\sin\alpha_h \right\rangle\end{pmatrix}, \quad (2)$$

where $\delta a_h^{(i)} = e\delta A_h^{(i)}/m_e c^2$ are the normalized amplitudes, $\omega_b^2 = 4\pi e^2 n_b / m_e c^2$ is the square of the beam plasma frequency, $m_e$ and $c$ are the electron rest mass and the speed of light in vacuum. The averaging operator is defined as

$$\langle(...)\rangle = \int_0^{2\pi} \frac{\sigma(\psi_0)}{2\pi} d\psi_0 \int_0^\infty dp\ G_0(p_z)(...)\ .$$

Here, $\sigma(\psi_0)$ is the phase distribution at entry time, and $\psi_0 = \omega t_0$ (where $t_0$ is the entry time) is the initial phase. Also, $G_0(p_z)$ is the initial momentum space distribution which is chosen to have a spread in the longitudinal momentum, in the form of a Gaussian distribution function, without any spread in the transverse momentum. The field equations are used in steady-state simulation. The planar wiggler in one dimension is written as

$$\mathbf{B}_w(z) = B_w(z)\sin(k_w z)\hat{e}_y, \qquad (3)$$

For entrance into the interaction region, an adiabatic entry field of the wiggler is necessary in order to inject electrons into the equilibrium and steady-state trajectories. The adiabatic magnetic field is

$$B_w(z) = \begin{cases} B_w \sin^2\left(\dfrac{k_w z}{4N_w}\right), & 0 \leq z \leq N_w \lambda_w \\ B_w, & N_w \lambda_w < z \end{cases}.$$

By using dimensionless variables $\mathbf{u} = \mathbf{p}/m_e c$, $\bar{k}_0 = k_0/k_w$, and $\bar{z} = k_w z$, beam dynamic equations, using the Lorentz force equations, are

$$\begin{aligned}\frac{du_x}{d\bar{z}} &= \overline{\Omega}_n(\bar{z})\sin(\bar{z}) + \sum_{h,odd}^n h(\bar{k}_0 - \frac{\gamma\overline{\omega}}{u_z}) \\ &\times (\delta a_h^{(1)}\cos\alpha_h - \delta a_h^{(2)}\sin\alpha_h),\end{aligned} \qquad (4)$$

$$\frac{du_y}{d\bar{z}} = 0, \tag{5}$$

$$\frac{du_x}{d\bar{z}} = -\bar{\Omega}_n(\bar{z})\sin(\bar{z})\frac{u_x}{u_z} - \frac{u_x}{u_z}\sum_{h,odd}^{n} h\bar{k}_0$$
$$\times (\delta a_h^{(1)}\cos\alpha_h - \delta a_h^{(2)}\sin\alpha_h), \tag{6}$$

$$\frac{d\alpha_h}{d\bar{z}} = h\left(\bar{k}_0 - \frac{\bar{\omega}\gamma}{u_z}\right), \tag{7}$$

where $\bar{\Omega}_n = eB_{w,n}/m_e k_w c^2$ and index $n$ in $B_{w,n}$ shows the wiggler magnetic field intensity at different settings.

In this paper, we consider a harmonic lasing FEL in which the wiggler consists of two halves with two different magnetic field strengths but the same wavelength $\lambda_w$. We reduce the wavelength of the fundamental harmonic of the first part by reducing the rms wiggler $a_n = \bar{\Omega}_n/\sqrt{2}$ (for planar wiggler). Suppose that in the first part the rms wiggler parameter is $a_1$ and the fundamental resonant wavelength is $\lambda_1$ with the harmonic resonant wavelength $\lambda_h = \lambda_1/h$, $h = 3,5,7,...$ . In the second part the rms wiggler parameter is $a_n$ so that the new resonant fundamental wavelength is the $n$th harmonic of the first mode setting, $\lambda_1' = \lambda_n$. We assume that the beam energy and the undulator period are fixed, Therefore the retuned wiggler parameter $a_n$ is obtained from the FEL resonance relation:

$$\frac{1+a_1^2}{1+a_n^2} = n. \tag{8}$$

Obviously, there are no real solutions for $a_n$ if $a_1 < a_c = \sqrt{n-1}$. As a result, $a_1$ must be larger than $a_c$. In this configuration, the $n$th harmonic radiation of the first part of the wiggler coincides with the fundamental radiation of the second part and therefore it grows to saturation whereas the fundamental wavelength of the first part is nonresonant and therefore its growth will be disrupted in the second part.

According to Ref. 40, with the universal scaling notation, the normalized self-consistent equations, which can be used to describe the planar wiggler FEL in 1D Compton limit, are

$$\frac{d\theta_j}{d\bar{z}} = p_j, \tag{9}$$

$$\frac{dp_j}{d\bar{z}} = -\sum_{h,odd} F_h(\xi_w)(\bar{A}_h e^{ih\theta_j} + c.c.), \tag{10}$$

$$\frac{d\bar{A}_h}{d\bar{z}} = F_h(\xi_w)\langle e^{-ih\theta}\rangle, \tag{11}$$

where $j = 1,...,N$ are the number of electrons, $\xi_w = a_w^2 / 2(1 + a_w^2)$, $F_h(\xi_w) = J_{(h-1)/2}(h\xi_w) - J_{(h+1)/2}(h\xi_w)$, with $J_h$ being the Bessel function of the $h$th order of the first kind, $p_j \equiv (\gamma_j - \gamma)/\rho\gamma$, and $\rho \equiv (1/\gamma) \times [(a_w/4)(\omega_b/ck_w)]^{2/3}$ is the pierce parameter.

For the harmonic lasing interaction with the wiggler parameters $a_1$ and $a_n$ and wavelengths $\lambda_1$ and $\lambda_n$, using identical scaling as Eqs. (9)-(11) and neglecting all harmonics $h > n$, the FEL equations may be obtained as

$$\frac{d\theta_j}{d\bar{z}} = \frac{1}{n} p_j, \tag{12}$$

$$\frac{dp_j}{d\bar{z}} = -\frac{a_n}{a_1} F_1(\xi_n)(\bar{A}_n e^{in\theta_j} + c.c.), \tag{13}$$

$$\frac{d\bar{A}_n}{d\bar{z}} = F_1(\xi_n)\langle e^{-in\theta} \rangle. \tag{14}$$

The 1D model gives the highest possible FEL gain (shortest gain length) and can be used as a reference for the cases with real electron beam. For real beam, the 3D effects (emittance, diffraction, and energy spread) contribute to the increase of the gain length. We can include the 3D effects as modifications to the 1D model via Ming-Xie parameterization[41,42].

$$\chi_d = \frac{L_{1D}}{Z_R}, \quad \chi_\varepsilon = \frac{L_{1D}}{\beta_{ave}} \frac{4\pi\varepsilon_u}{\lambda_r}, \quad \chi_\gamma = \frac{4\pi L_{1D}}{\lambda_u} \frac{\delta\sigma_\gamma}{\gamma}$$

where $Z_R = \frac{4\pi\sigma_x^2}{\lambda} = \frac{\pi w_0^2}{\lambda}$ is the Reyleigh length and $L_{1D}$ is the gain length given by the 1D model. Real gain length for FEL can be expressed as

$$\frac{L_{1D}}{L_g} = F(\chi_d, \chi_\varepsilon, \chi_\gamma).$$

To minimize the 3D effects, Ming-Xie parameters should be less than 1.

### III. THE NUMERICAL ANALYSIS

The complete set of coupled nonlinear differential equations is solved numerically using the fourth-order Rounge-Kutta method. For the particle averaging, the Gaussian quadrature technique in each of the degrees of freedom $(\psi_0, p_{z0})$ is used. In this work, an attempt was made to match the third harmonic resonance of the first part of the wiggler to the fundamental resonance of the second part in order to obtain UV radiation. The common parameters of the wigglers, radiation, and the electron beam which are used in this study are as follows: the electron beam has the relativistic factor of 300 with peak current of 200A and the initial radius of 0.15 mm. The wavelength of both parts of the wiggler is 2.8 cm. The peak value of the on-axis amplitude of the first part of the wiggler is 8.33 KG over the length of 5 m, and the entry tapered region is $N_w = 10$ wiggler period in length. Using relation (8), we obtain $a_3 = 0.349$, and optimize the step taper and overall length for the maximum output. Using these beam and wiggler parameters, the fundamental resonance is at the wavelength of 523.9 nm, and the initial power is 10 W. The third harmonic resonance wavelength is 174.63 nm and starts from zero initial power. The saturation power and the saturation length of the third harmonic radiation depends on the length of the first part of the wiggler $L_1$. The power of the third harmonic lasing versus distance, for different lengths of the first part of the wiggler $L_1 = 1.5m$ (dashed-dotted line), $L_1 = 2m$ (dotted line), $L_1 = 4.5m$ (dashed line), $L_1 = 5m$ (solid line), is shown in Fig. 2. It can be seen that for $L_1 = 1.5m$ the power decreases in the beginning of the second part of the wiggler so the saturation power decreases and the saturation length increases. Because the electron beam in the first part of the wiggler were bunched according to the first harmonic. So the electron beam in the beginning of the second part of the wiggler, in which the third harmonic of the first part of the wiggler is a seed for the second part, is not uniformly distributed. This difference in the distribution of the electrons at the different lengths of the first part of the wiggler leads to changes in the saturation power and saturation length in the second part of the wiggler, compared to the conventional wiggler.

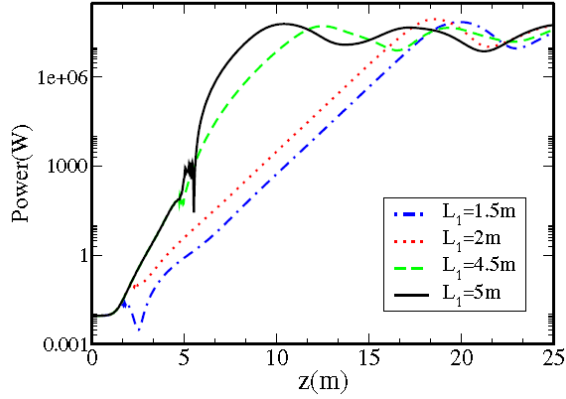

Fig. 2. Power of the third harmonic for $L_1 = 1.5m$, $L_1 = 2m$, $L_1 = 4.5m$, $L_1 = 5m$ versus z(m).

Figure 3 shows how these two quantities vary with $L_1$. This shows that by choosing $L_1$ around $5m$, which is approximately where the fundamental wavelength saturates in the first part of the wiggler, optimized saturation power and saturation length are obtained.

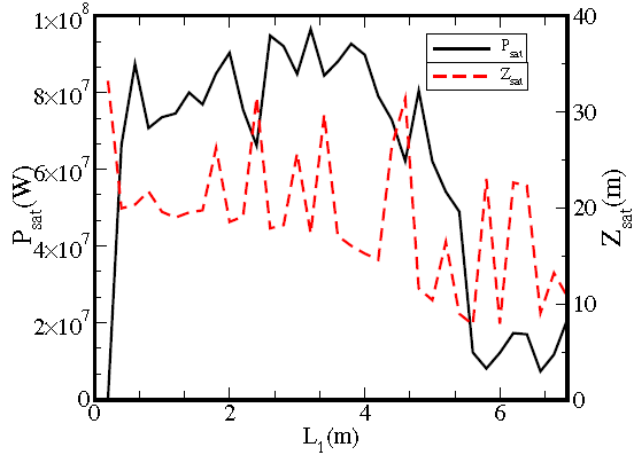

Fig. 3. Variation of saturated power (solid line) and saturation distance (dashed line) of the third harmonic versus the length of the first part of the wiggler.

Figure 4 demonstrates the harmonic lasing scheme as described above. The resonant, cold beam limit is assumed. It is seen that the fundamental scaled intensity (a) is disrupted at $z = 5m$ with $\overline{\Omega}_1 = 2.18$. Also, the intensity of the third harmonic (b), which is the fundamental of the second part of the wiggler when the undulator parameter is re-tuned at $\overline{\Omega}_3 = 0.49$, and the intensity of the fifth harmonic(c) are shown in Fig. 4. It is seen that the third harmonic grows to $3.8 \times 10^7 W$ and the fifth harmonic is also disrupted.

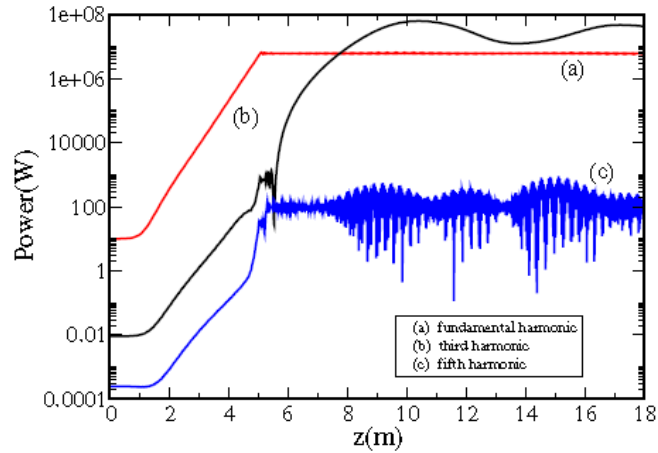

Fig. 4. Power of the fundamental resonance (a), the third harmonic (b), and the fifth harmonic (c) versus $z(m)$.

## A: COMPARISON BETWEEN NONAVERAGED AND AVERAGED CODES

Equations (2),(4)-(7) are obtained with the non-averaged treatment and Eqs. (9)-(14) that are averaged on the undulator period have been solved numerically with two independent codes.

The simulations with non-averaged and averaged equations are compared in one dimension. A steady-state regime is considered in which the pulse length is assumed to be much longer than the slippage length over the entire wiggler. Both codes use the quite start and the beam is assumed to be cold.

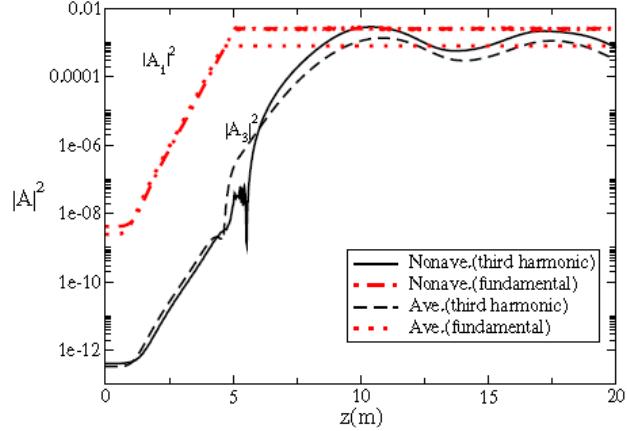

Fig. 5. Comparison between non-averaged and averaged model for $|A_1|^2$ of the fundamental and $|A_3|^2$ of the third harmonic.

Figure 5 shows the evolution of the radiation intensity, $|A|^2$, for the fundamental and the third harmonic by the two simulation codes. As can be seen, in both cases, the third harmonic saturates at $z = 10.4 m$ and both have the same growth rate and gain length. However, field intensity in the average code at $|A|^2{}_{ave} = 0.0014$ is less than that of the non-average code at $|A|^2{}_{nonave} = 0.0027$. The difference of field intensities between the two codes may be the result of differences in the set of equations and the integration methods.

## B: ENERGY SPREAD EFFECTS

Higher harmonics are more sensitive to the energy spread than the fundamental one.[4,18] The work of Ref. 18 showed that the third harmonic lasing is much more sensitive to the effects of the electron beam energy spread than the retuned fundamental lasing.

To consider effects of the energy spread, we choose the initial conditions that model the axial injection of the electron beam with energy in the form of a Gaussian distribution function that is peaked around the initial energy of the beam. We chose the thermal distribution function as

$$G_0(p_z) = \sqrt{\frac{2}{\pi}} \frac{1}{\Delta p_z} \exp\left(-\frac{2(p_z - p_0)^2}{\Delta p_z}\right), \tag{15}$$

where $p_0$ and $\Delta p_z$ are the initial bulk momentum and momentum spread, respectively. The corresponding axial energy spread can be written as

$$\frac{\Delta \gamma_{z0}}{\gamma_0} = \frac{\gamma_0^2 - 1}{\gamma_0^2} \frac{\Delta p_{z0}}{p_0}, \tag{16}$$

where $\gamma = \sqrt{1 + \frac{p_0^2}{m^2 c^2}}$ is the average initial energy of the electron at $t_0$ (injection time).

The n-fold increase in the phase velocity spread for harmonic lasing also has the effect of increasing the homogeneous energy requirements of the electron beam at the beginning of the interaction. This may be expressed as $\sigma_\gamma/\gamma \langle \rho$ or equivalently as $\sigma_p \langle 1$. The final power is reached after a series of oscillations due to the phenomenon of mutual bunching as described in Ref. 43. This requirement is increased to $\sigma_p \langle 1/n$ or $\sigma_\gamma/\gamma \langle \rho/n$ for harmonic lasing. Figure 6 shows $\sigma_\gamma/\gamma$, obtained from the phase space, versus z along the wiggler for a conventional FEL (a) and harmonic lasing (b). This effect of energy spread on harmonic lasing is evident by comparing plots (a) and (b).

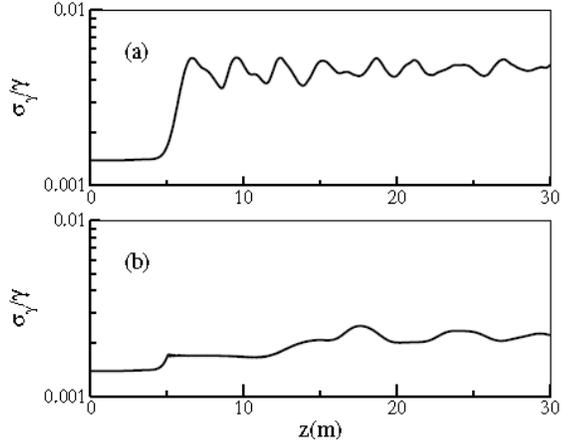

Fig. 6. The relative quantity $\sigma_\gamma/\gamma$, computed by the phase space, versus $z$ along the undulator.

It is interesting to study how the initial thermal distribution function $G_0$ of the electron beam, as given by Eq.(15), changes as the beam approaches the saturation point. To find $G$ at different points along the wiggler a histogram is drawn by counting the number of electrons in small intervals of $\gamma$. If these intervals of $\gamma$ are small enough, a continuous curve will be obtained. In Fig. 7, the initial thermal distribution function $G_0$ (dashed line) is compared with that at the saturation point $G_s$ (solid line) of the third harmonic. For a given energy spread with $\Delta\gamma/\gamma = 0.002$, the initial distribution function $G_0$ that is at the interval $299 < \gamma < 301$ is further thermalized at the saturation point around $298 < \gamma < 301$. It can be seen that the initial distribution function is shifted to lower energies at saturation. This shows that the electrons have lost their kinetic energy in order to amplify the radiation. Areas under the distribution function that represent the total number of electrons are normalized to unity. The fact that the initial energy distribution function of electrons $G_0$ deviates from the Gaussian distribution is the artifact of the Gauss-quadrature algorithm and is compensated for by non-uniform weights assigned to each electron.

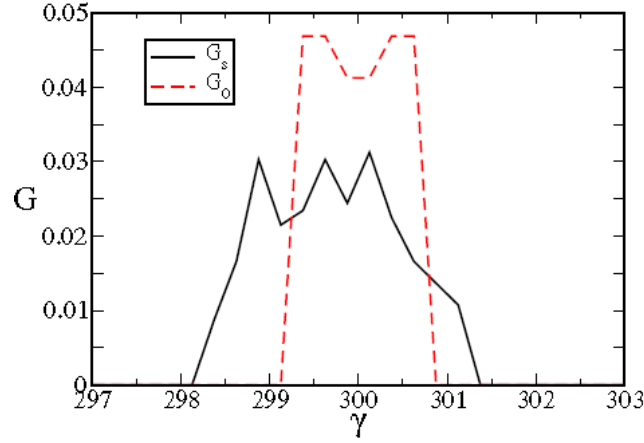

Fig. 7. Energy distribution function of the electron beam with $\Delta\gamma/\gamma = 0.002$.

Figure 8 shows the longitudinal phase space of the electrons at different coordinates $z$ along the wiggler for $\Delta\gamma/\gamma = 0.002$. Plots (a),(b), and (c) are for the first part of the wiggler and plots (d), (e), and (f) are related to the second part. The uneven distribution of initial phase, in plot (a), is the artifact of the Gauss quadrature weightings. Observe that as electrons move along the wiggler their energy gets redistributed in pondermotive phase $\psi = \alpha_3 + k_w z$ along a straight line with a sinusoidal modulation. This straight line is an indication of the energy spread. It is evident that the spread of energy becomes progressively larger as the beam approaches the saturation point. At $z = 6m$, the sinusoidal modulation is an indication that the untrapped electrons move along and pass over the pondermotive potential. The trapped particle at saturation is shown by the non-sinusoidal phase-space plot at $z = 17m$. We can conclude, therefore, that the plot of $\gamma$ versus $\psi$ truly represents the conventional phase space of $d\psi/dz$ versus $\psi$.

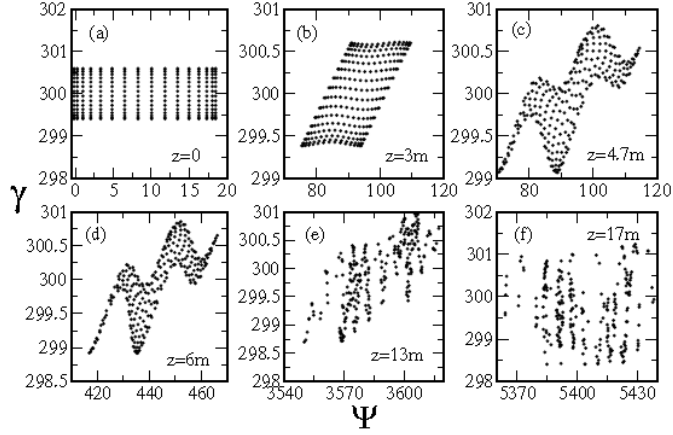

Fig. 8. Distribution of energy $\gamma$ versus the distribution of pondermotive phase of the electron beam at different values of the coordinate $z$ along the undulator.

Thermalization and trapping of electrons in the course of their motion along the wiggler, as indicated in Fig. 8, might show some sign of chaoticity. This may be investigated using the Kolmogorov entropy. The sensitivity of the motion of an electron to small changes in initial conditions may be demonstrated using the Kolmogorov entropy.[44,45] If the close neighbor trajectories separate linearly in time the motion is regular, while the motion is chaotic if they diverge exponentially. We can define the Kolmogorov entropy as:

$$k_N = \frac{1}{\tau N} \sum_1^N \ln \frac{|d_i|}{|d_0|}, \qquad (17)$$

where N is the number of small time steps $\tau$, $N\tau$ is equal to the total time $T$ of the motion, $|d_i|$ is the norm of the separation of the two trajectories in the phase space with six dimensions, and $|d_0|$ is its initial value.[43] Divergence of the trajectories of two neighboring electrons from each other is determined by the value of $k_N$. For small separations of the trajectories, $k_N$ is small. When $k_N$ is positive the motion is chaotic and close neighbor trajectories separate exponentially.

When $k_N$ is negative, however, the motion is regular and the neighboring trajectories approach each other.

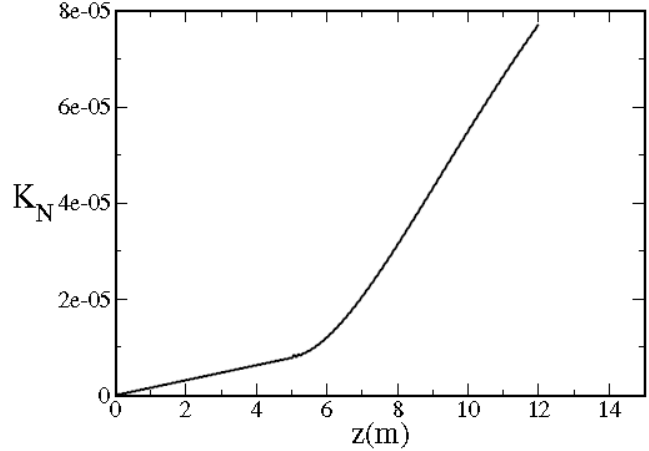

Fig. 9. The evolution of $k_N$ with $z$ in the absence of radiation.

The evolution of $k_N$ with $z$ in the absence of radiation is shown in Fig. 9. The iteration step is $597.4\,nm$, and the total iteration number is $2\times 10^7$. The reference electron is initially located at $x = 0.0065\,mm$ with $y = z = 0$. The relativistic factor of the reference electron is 300. The neighboring electron is at the same position with a slightly different relativistic factor $300\times$ $(1+10^{-8})$. The increasing positive values of $k_N$ in Fig.9 show that the dynamical stability decreases by increasing the Kolmogorov entropy.

**C: TAPERED WIGGLER IN HARMONIC LASING**

The radiation intensity, $|A|$, indicates the efficiency of the FEL system. The lower(higher) intensity yields a lower(higher) efficiency.

In the absence of tapering, oscillation of the radiation power beyond the saturation point persists. In one period, electrons give their energy to the radiation to increase the radiation amplitude and reduce the energy of electrons. This will cause the electrons to lose their resonance with the electromagnetic radiation and go to the phase in which they will extract energy from the radiation. Following this situation, the amplitude of radiation decreases up to the point at which electrons gain energy and their resonance with the radiation is established. This will cause the electrons to lose their energy to radiation. This cycle will repeat itself. It is possible to increase the efficiency by decreasing the phase velocity of the radiation or accelerating the electrons so that the reduction in the electron beam energy is compensated. It has been shown that by decreasing the amplitude of the wiggler a portion of the transverse energy of the electron beam will be transferred to the longitudinal motion and therefore accelerate the electron beam. This reduction can be linearly or linearly, that nonlinear decreasing can obtain more suitable restoration of the resonance condition.

It is assumed that the FEL has a constant wiggler field $B_w$ (beyond the injection region) up to the point $\bar{z}_T$ (the saturation point), and after that the wiggler amplitude decreases linearly by the slope $m$. The linearly tapered wiggler may be written as

$$\overline{\Omega}_{wh}(\bar{z}) = \begin{cases} \overline{\Omega}_{wh}, & \bar{z} < \bar{z}_T \\ \overline{\Omega}_{wh} - m(\bar{z} - \bar{z}_T), & \bar{z} \geq \bar{z}_T \end{cases}. \qquad (18)$$

And the nonlinearly tapered wiggler can be as

$$\overline{\Omega}_{wh}(\overline{z}) = \begin{cases} \overline{\Omega}_{wh}, & \overline{z} < \overline{z}_{T1} \\ \overline{\Omega}_{wh} - m_1(\overline{z} - \overline{z}_{T1}), & \overline{z}_{T1} \leq \overline{z} \leq \overline{z}_{T2} \\ \overline{\Omega}_{wh}(z_{T2}) - m_2(\overline{z} - \overline{z}_{T2}), & \overline{z}_{T2} \leq \overline{z} \leq \overline{z}_{T3} \\ \vdots \\ \overline{\Omega}_{wh}(z_{T(n-1)}) - m_{(n-1)}(\overline{z} - \overline{z}_{T(n-1)}), & \overline{z}_{T(n-1)} \leq \overline{z} \end{cases} \quad (19)$$

$m_i$ with $i = 1, 2, ...$ are slopes of the lines between $z_{Ti}$ and $z_{Ti+1}$.

The second part of the wiggler in a tapered harmonic lasing has been optimized over the entire length of 16.1 m. In Fig. 10(a), a comparison is made between the linearly tapered (dashed line), nonlinearly tapered (solid line), and untapered (dotted line) wigglers for the power of the third harmonic lasing. The wiggler amplitude is decreased at $z_T = 16.1 m$ with the slope $m = 0.2 \times 10^{-4}$ for the linearly tapered wiggler. For nonlinear tapering, the amplitude of the wiggler is decreased at $z_{T1} = 16.1m$, $z_{T2} = 19.0m$, $z_{T3} = 22.3m$, $z_{T4} = 25.5m$, $z_{T5} = 28.3m$, $z_{T6} = 30.7m$ with slopes of $m_1 = 0.2 \times 10^{-4}$, $m_2 = 0.3 \times 10^{-4}$, $m_3 = 0.6 \times 10^{-4}$, $m_4 = 0.8 \times 10^{-4}$, $m_5 = 0.9 \times 10^{-4}$, $m_6 = 1.2 \times 10^{-4}$, respectively. The parameters $m$ and $z_T$ are chosen to obtain the maximum intensity of the third harmonic lasing, which are found by a successive run of the code. We can see significant increases in the third harmonic power from $3.8 \times 10^7$ to $3.5 \times 10^8$ W for the linear tapering, and from $3.8 \times 10^7$ to $8.1 \times 10^8$ W for the nonlinear tapering. It can be seen that the nonlinear tapering of the wiggler amplitude with different slopes has higher efficiency. The profile for the variation of the wiggler parameter $\overline{\Omega}_w$ with $z(m)$ for the nonlinear tapering (solid line) and linear tapering (dashed line) of the wiggler is shown in Fig. 10(b).

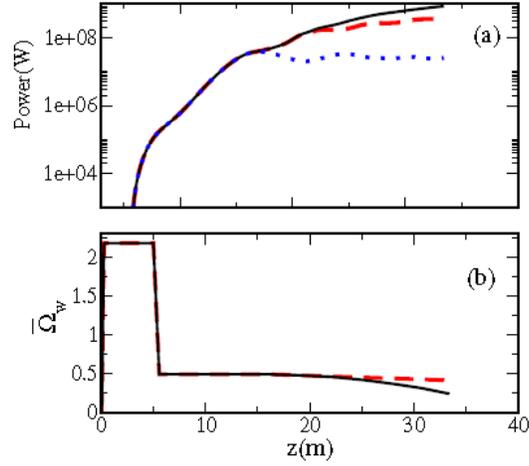

Fig. 10. Comparison of the growth in the third harmonic power for a linearly tapered (dashed line), nonlinearly tapered (solid line), and untapered (dotted line) wiggler(a). Plot (b) shows $\bar{\Omega}_w$ versus $z$ for the linear tapering (dashed line) and nonlinear tapering (solid line) of the wiggler.

The longitudinal phase space of the electrons at different coordinates $z$ along the linearly tapered wiggler is represented in Fig. 11. It can be seen that in the tapered region and beyond the energy of the electron beam is reduced. Because of the restoration of the resonance condition, electrons give their energy to the radiation and the radiation amplitude increases.

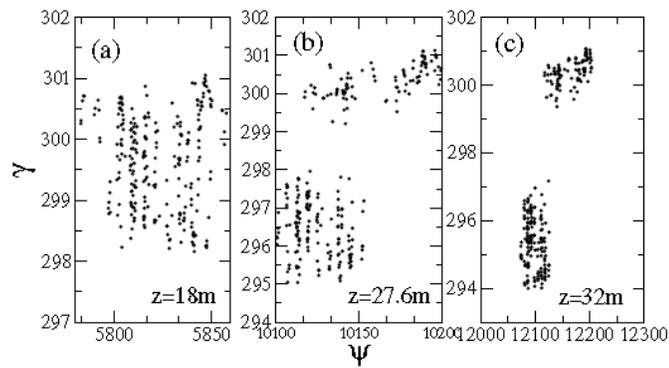

Fig. 11. Distribution of energy $\gamma$ versus the distribution of pondermotive phase of the electron beam at different values of the coordinate $z$ along the linearly tapered wiggler.

In the short wavelength FEL, which needs to be in amplifier mode, the wiggler is usually too long. One can use a prebunched beam instead of the uniform one to reduce the length of the wiggler. We will show that prebunching for harmonic lasing, in the absence of tapering, increases the saturation power and reduces the wiggler length. The superposition of radiation fields emitted by electrons is initially zero for an unbunched beam. For a prebunched beam, however, the distribution of initial phases of the beam's electrons deviates from uniformity and the superposition of the fields will be nonzero. Therefore, a significant radiation field at the entrance to the wiggler builds up. When an unbunched electron beam is used in a FEL the initial external seed forces the electron beam to get bunched during the first few wiggler periods due to the pondermotive potential formed by the beating of the wiggler and the radiation field. Since this bunching exists at the entrance to the wiggler, in a prebunched FEL, the initial external seed radiation is not necessary. Figure 12 compares the powers of the third harmonic for a uniform and a prebunched beam. It is evident that the saturation length is reduced and the power is increased.

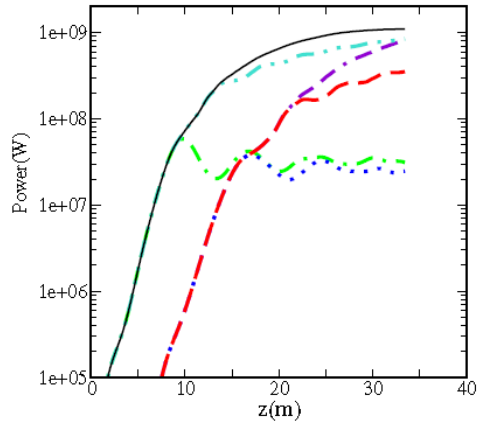

Fig. 12. Comparison of the growth in third harmonic powerfor a uniform (dotted line) and a prebunchedbeam (dashed-dotted line) without tapered wiggler and for a uniform (dashed line) and a prebunched beam (dashed-double

dotted line) with linearly tapered wiggler with $z_T = 9.7m$ and $m = 0.4 \times 10^{-4}$. The same comparison for a uniform (double dashed-dotted line) and a prebunched beam (solid line) with nonlinearly tapered wiggler.

In Fig. 12, for the uniform beam, the dotted line shows that the radiation of the third harmonic saturates at $z = 17.05m$ with the saturated power of $3.8 \times 10^7 W$ and for the prebunched beam with $\psi_{width} = 1.85\pi$, the radiation of the third harmonic saturates at $z = 9.8m$ with the saturated power of $5.8 \times 10^7 W$ (dashed-dotted line). We can see in Fig. 12 that with the linearly tapered wiggler, for the uniform beam (dashed line), the radiation of the third harmonic saturates at $z = 33.4m$ with the saturated power of $3.5 \times 10^8 W$. For the prebunched beam with $\psi_{width} = 1.85\pi$ and linear tapering, the dashed-double dotted line in Fig. 12 shows that the third harmonic saturates at $z = 33.4m$ with the saturated power of $8.5 \times 10^8 W$. The length of the first wiggler is $L_1 = 0.6m$ and the tapering starts at $z_T = 8.6m$ with the slope $m = 0.4 \times 10^{-4}$. When we decrease the amplitude of the wiggler nonlinearly, in Fig. 12 for a uniform beam (dashed-double dotted line), the saturated power of the third harmonic is $8.1 \times 10^8 W$ at $z = 33.4$ and for the prebuched beam with $\psi_{width} = 1.85\pi$ (solid line), the third harmonic saturates at $z = 33.01$ with the saturated power of $1.1 \times 10^9 W$, which the wiggler amplitude is reduced at $z_{T1} = 8.6$, $z_{T2} = 11.3$, $z_{T3} = 13.7$, $z_{T4} = 17.5$, $z_{T5} = 21.3$ (all in meters) with slopes of $m_1 = 0.4 \times 10^{-4}$, $m_2 = 0.5 \times 10^{-4}$, $m_3 = 0.8 \times 10^{-4}$, $m_4 = 1.0 \times 10^{-4}$, $m_5 = 1.1 \times 10^{-4}$, respectively. As it is seen in Fig. 12, the saturation length for the uniform beam (dashed line) is the same as for the prebunched beam (solid line).

## IV. CONCLUSION

In this paper, a one-dimensional simulation is conducted to analyze harmonic lasing FEL. In order to generate UV radiation, the fundamental resonance disrupted by reducing the undulator magnetic field and the third harmonic grows up to the saturation point. Non-averaged and averaged (MEDUSA) equations have been written and integrated. The thermal effect of the electron beam is taken into account. By relating the energy distribution function to the distribution function of the pondermotive phase, chaotic patterns at the saturation point are revealed. This is investigated by computing the Kolmogorov entropy. It is found that by suitably tapering the amplitude of the second wiggler field, saturation of the radiation for a shorter wavelength can be postponed leading to further amplification. In the case of the tapered wiggler, it is found that the efficiency can be enhanced more effectively by the nonlinear reduction of the second wiggler field at saturation point compared to that of the linear tapering. Also, prebunching the beam yields substantial advantages for the harmonic lasing FEL operation and permits higher saturation power and shorter saturation lengths.